\documentclass{ws-p10x7}

\begin{document}

\title{Flavor Symmetry and Neutrino Oscillations}

\author{Yue-Liang Wu}

\address{Institute of Theoretical Physics, \\ Chinese Academy of Sciences, Beijing 100080, China \\
E-mail: ylwu@itp.ac.cn}

\twocolumn[\maketitle\abstract{ We show how the nearly bi-maximal mixing 
scenario comes out naturally from gauged $SO(3)_{F}$ flavor symmetry via
spontaneous symmetry breaking. An interesting relation between the neutrino
 mass-squared differences and the mixing angle, i.e., 
$\Delta m_{\mu e}^2 / \Delta m_{\tau\mu}^2 \simeq 2|U_{e 3}|^2 $  is obtained. 
The smallness of the ratio (or $|U_{e 3}|$) can also naturally be understood 
from an approximate permutation symmetry. Once the mixing element $|U_{e 3}|$ is determined, 
such a relation will tell us which solution will be favored within this model.
The model can also lead to interesting phenomena on lepton-flavor violations. 
}]

 \section{Introduction} 
The greatest success of the  standard model (SM) is the gauge symmetry structure 
$SU(3)_{c}\times SU_{L}(2)\times U_{Y}(1)$ which has been tested by more and 
more precise experiments. In the SM, neutrinos are assumed to be massless. 
The recent evidences for oscillation of atmospheric neutrinos\cite{SK1} 
and for the deficit of the measured solar neutrino flux\cite{SK2} strongly suggest that  
neutrinos are massive though their masses are small, and new physics beyond the SM is necessary. 
The scenario most favoured by the current data\cite{SK} may comprise just three light neutrinos with 
nearly bimaximal mixing via MSW solution\cite{MSW}. 
It is of interest to note that such a scenario 
was shown to be naturally obtained from a simple extention of the SM with gauged SO(3)$_{F}$ flavor 
symmetry\cite{YLW}. In this talk I mainly describe the most interesting features 
resulting from such a simply extended model.  

\section{The model}

 For a less model-dependent analysis, we directly start from an $SO(3)_{F}\times 
SU(2)_{L}\times U(1)_{Y}$ invariant effective lagrangian with three $SO(3)_{F}$ Higgs triplets 
\begin{eqnarray}
& & {\cal L} =  \frac{1}{2}g'_{3}A_{\mu}^{k}
( \bar{L}_{i}\gamma^{\mu} (t^{k})_{ij}L_{j} 
+ \bar{e}_{R i} \gamma^{\mu}(t^{k})_{ij}e_{R j} ) \nonumber \\
& & + ( Y_{1 ij} \bar{L}_{i} \phi_{1}e_{R\ j} + 
Y_{2 ij}\bar{L}_{i} \phi_{2}\phi_{2}^{T}L_{j}^{c} + H.c. ) \nonumber \\
& & + D_{\mu}\varphi^{\ast} D^{\mu}\varphi + D_{\mu}\varphi^{'\ast}D^{\mu}\varphi'
  + D_{\mu}\varphi^{''\ast} D^{\mu}\varphi'' \nonumber \\
& & - V_{\varphi}  + {\cal L}_{SM}   
\end{eqnarray}
with effective Yukawa couplings
\begin{eqnarray}
 Y_{1 ij} & = & c_{1}\varphi_{i}\varphi_{j}\chi +  c'_{1} \varphi'_{i}\varphi'_{j}\chi' + 
c''_{1} \varphi''_{i}\varphi''_{j}\chi'' \nonumber \\
Y_{2 ij} & = & c_{0}\varphi_{i}\varphi_{j}^{\ast} + c'_{0}\varphi'_{i}\varphi_{j}^{'\ast} 
+ c''_{0}\varphi''_{i}\varphi_{j}^{''\ast} + c\delta_{ij}  \nonumber 
\end{eqnarray} 
${\cal L}_{SM} $ denotes the lagrangian of the standard 
model. $\bar{L}_{i}(x) = (\bar{\nu}_{i}, \bar{e}_{i})_{L}$ 
(i=1,2,3) are the SU(2)$_{L}$ doublet leptons and $e_{R\ i}$ ($i=1,2,3$) are 
the three right-handed charged leptons. $A_{\mu}^{i}(x)t^{i}$ ($i=1,2,3$) 
are the $SO(3)_{F}$ gauge bosons with $t^{i}$ the $SO(3)_{F}$ generators and 
$g'_{3}$ is the corresponding gauge coupling constant. Here $\phi_{1}(x)$ and $\phi_{2}(x)$ 
are two Higgs doublets, $\varphi(x)$, $\varphi'(x)$ and $\varphi''(x)$ are
three $SO(3)_{F}$ Higgs triplets, and $\chi(x)$, 
$\chi'(x)$ and $\chi''(x)$ are three singlet scalars. The couplings $c$, $c_{a}$, $c'_{a}$ and 
$c''_{a}$ ($a=0,1$) are dimensional constants. The structure of the above effective 
lagrangian can be obtained by imposing an additional U(1) symmetry \cite{YLW}. 

 The Higgs potential for the $SO(3)_{F}$ Higgs triplets has the following general form before 
symmetry breaking 
\begin{eqnarray}
& & V_{\varphi}  =  \frac{1}{2} \mu^{2} (\varphi^{\dagger}\varphi ) + 
\frac{1}{2} \mu^{'2} (\varphi'^{\dagger}\varphi' ) + 
\frac{1}{2} \mu^{''2} (\varphi''^{\dagger}\varphi'' )  \nonumber \\
& & +  \frac{1}{4}\lambda (\varphi^{\dagger}\varphi)^{2} + 
\frac{1}{4}\lambda' (\varphi'^{\dagger}\varphi')^{2} + 
\frac{1}{4}\lambda'' (\varphi''^{\dagger}\varphi'')^{2} \nonumber \\
& & + \frac{1}{2}\kappa_{1} (\varphi^{\dagger}\varphi )( \varphi'^{\dagger}\varphi') + 
\frac{1}{2}\kappa'_{1} (\varphi^{\dagger}\varphi )( \varphi''^{\dagger}\varphi'') \nonumber  \\
& & + \frac{1}{2}\kappa''_{1} (\varphi'^{\dagger}\varphi' )( \varphi''^{\dagger}\varphi'') 
 + \frac{1}{2} \kappa_{2} (\varphi^{\dagger}\varphi')( \varphi'^{\dagger}\varphi ) \nonumber \\
& & +  \frac{1}{2} \kappa'_{2} (\varphi^{\dagger}\varphi'')( \varphi''^{\dagger}\varphi )
+ \frac{1}{2} \kappa''_{2} (\varphi'^{\dagger}\varphi'')( \varphi''^{\dagger}\varphi' )\  .  \nonumber 
\end{eqnarray}
 
   As the $SO(3)_{F}$ flavor symmetry is treated to be a gauge symmetry, 
one can always express the complex $SO(3)_{F}$ Higgs triplet  
field in terms of three rotational fields $\eta_{i}(x)$  and three 
amplitude fields $\rho_{i}(x)$ 
\begin{equation} 
 \left( \begin{array}{c}
  \varphi_{1}(x) \\
  \varphi_{2}(x)   \\
  \varphi_{3}(x)   \\
\end{array} \right) = e^{i\eta_{i}(x)t^{i}} \frac{1}{\sqrt{2}}
\left( \begin{array}{c}
  \rho_{1}(x) \\
  i\rho_{2}(x)   \\
  \rho_{3}(x)   \\
\end{array} \right) 
\end{equation}
Similar forms are for $\varphi'(x)$ and  $\varphi''(x)$. Assuming that
only the amplitude fields get VEVs after spontaneous symmetry breaking, 
namely  $<\rho_{i}(x)> = \sigma_{i}$, $<\rho'_{i}(x)> = \sigma'_{i}$ and  
$<\rho''_{i}(x)> = \sigma''_{i}$,  we then obtain the following equations from minimizing the 
Higgs potential\cite{YLW}  
\begin{eqnarray}
& & \sigma'_{1}= \sqrt{\xi} \sigma_{1}, \    \sigma'_{2}= \sqrt{\xi} \sigma_{2}, 
\   \sigma'_{3}= -\sqrt{\xi} \sigma_{3}, \nonumber \\ 
& & \sigma''_{1}= \sqrt{2\xi'} \sigma_{1}, \  \  \sigma''_{2}= -\sqrt{2\xi'} \sigma_{2}, 
\  \  \sigma''_{3}= 0, \nonumber  \\ 
& &  \sigma_{3}^{2} = \sigma_{1}^{2} + \sigma_{2}^{2} =2\sigma_{1}^{2}= \sigma^{2}/2 \ .
\end{eqnarray}
where we have asummed  a global minimum potential energy $V_{\varphi}|_{min}$ for 
varying $\xi$ and $\xi'$ at the minimizing point
\begin{eqnarray}
V_{\varphi}|_{min} & = & -\sigma^{4}(\lambda + \lambda'\xi^{2} + \lambda''\xi^{'2} \nonumber \\ 
& + & 2\kappa_{1}\xi + 2\kappa'_{1}\xi' + 2\kappa''_{1}\xi\xi')/4
\end{eqnarray}
with $\sigma^{2} = \sigma_1^2 + \sigma_2^2 + \sigma_3^2$, $\xi = \sigma^2/\sigma^{'2}$ 
and $\xi' = \sigma^2/\sigma^{''2}$.
It is seen that with these considerations the VEVs are completely determined 
by the Higgs potential.

  It is remarkable that with these relations the mass matrices of the neutrinos and 
charged leptons are simply given by 
\begin{eqnarray}
& & M_{e}  =  \frac{m_{\tau}}{2}\left( \begin{array}{ccc}
  \frac{1}{2} & \frac{1}{2}i & \frac{1}{\sqrt{2}}  \\
   \frac{1}{2}i & -\frac{1}{2} &  \frac{1}{\sqrt{2}}i \\
  \frac{1}{\sqrt{2}} & \frac{1}{\sqrt{2}}i & 1 \\ 
\end{array} \right)   \\
& &  +  \frac{m_{\mu}}{2}\left( \begin{array}{ccc}
 \frac{1}{2} & \frac{1}{2}i & -\frac{1}{\sqrt{2}}  \\
   \frac{1}{2}i & -\frac{1}{2} &  -\frac{1}{\sqrt{2}}i \\
  -\frac{1}{\sqrt{2}} & -\frac{1}{\sqrt{2}}i & 1 \\ 
\end{array} \right) - \frac{m_{e}}{2} \left( \begin{array}{ccc}
  1 & i & 0  \\
  i & -1  & 0 \\
 0 & 0 &  0 \\ 
\end{array} \right) \nonumber  \\
& & M_{\nu} = \hat{m}_{\nu}\left( \begin{array}{ccc}
  1  & 0 & \frac{1}{\sqrt{2}}\hat{\delta}_{-}  \\
  0 & 1   &  0 \\
  \frac{1}{\sqrt{2}}\hat{\delta}_{-} & 0
 & 1 + \hat{\Delta}_{-}  \\ 
\end{array} \right) 
\end{eqnarray}

\section{Nearly Bimaximal Mixing}

  It is more remarkable that the mass matrix $M_{e}$ can be diagonalized by a unitary 
bi-maximal mixing matrix $U_{e}$ via $D_{e} = U_{e}^{\dagger} M_{e} U_{e}^{\ast}$ with
\begin{equation}
U_{e}^{\dagger}=\left( \begin{array}{ccc}
  \frac{1}{\sqrt{2}}i & -\frac{1}{\sqrt{2}} & 0  \\
 \frac{1}{2} & -\frac{1}{2}i & -\frac{1}{\sqrt{2}} \\
 \frac{1}{2} & -\frac{1}{2}i & \frac{1}{\sqrt{2}}  \\
\end{array} \right)
\end{equation}
and $D_{e} = diag. (m_{e}, m_{\mu}, m_{\tau})$. 
 The neutrino mass matrix can be easily diagonalized 
by an orthogonal matrix $O_{\nu}$ via $O_{\nu}^{T}M_{\nu}O_{\nu}$ with 
$(O_{\nu})_{13}=\sin \theta_{\nu}\equiv  s_{\nu}$ and 
$\tan2\theta_{\nu} = \sqrt{2}\hat{\delta}_{-}/\hat{\Delta}_{-}$.
Thus the CKM-type lepton mixing matrix $U_{LEP}$ that   
appears in the interaction term 
${\cal L}_{W} = \bar{e}_{L}\gamma^{\mu}U_{LEP} \nu_{L} W_{\mu}^{-} + H.c. $
is given by $U_{LEP} = U_{e}^{\dagger}O_{\nu}$. Explicitly, one has 
\begin{equation}
U_{LEP}  = \left( \begin{array}{ccc}
  \frac{1}{\sqrt{2}}ic_{\nu} & -\frac{1}{\sqrt{2}} & \frac{1}{\sqrt{2}}is_{\nu}  \\
 \frac{1}{2}c_{\nu}+ \frac{1}{\sqrt{2}}s_{\nu} & -\frac{1}{2}i & 
\frac{1}{2}s_{\nu}-\frac{1}{\sqrt{2}}c_{\nu} \\
 \frac{1}{2}c_{\nu}-\frac{1}{\sqrt{2}}s_{\nu} & -\frac{1}{2}i & 
\frac{1}{2}s_{\nu}+ \frac{1}{\sqrt{2}}c_{\nu}  \\
\end{array} \right).
\end{equation}
The three neutrino masses are found to be
\begin{eqnarray}
m_{\nu_{e}} & = &  \hat{m}_{\nu}[1 -  
(\sqrt{\hat{\Delta}_{-}^{2} + 2 \hat{\delta}_{-}^{2}} - \hat{\Delta}_{-} )/2\  ]
\nonumber \\
m_{\nu_{\mu}} & = &  \hat{m}_{\nu}   \\
m_{\nu_{\tau}} & = & \hat{m}_{\nu}[1 + \hat{\Delta}_{-} + 
(\sqrt{\hat{\Delta}_{-}^{2} + 2 \hat{\delta}_{-}^{2}} - \hat{\Delta}_{-} )/2\  ] . \nonumber 
\end{eqnarray}

  The similarity between the Higgs triplets $\varphi(x)$ and $\varphi'(x)$
naturally motivates us to consider an approximate (and softly broken) permutation 
symmetry between them. This implies that 
 $|\hat{\delta}_{-}|<< 1$. To a good approximation, the mass-squared 
differences are given by  $\Delta m_{\mu e}^{2} \equiv m_{\nu_{\mu}}^{2} - 
m_{\nu_{e}}^{2} \simeq  \hat{m}_{\nu}^{2}\hat{\Delta}_{-}(\hat{\delta}_{-}/\hat{\Delta}_{-})^{2}$ 
and  $\Delta m_{\tau\mu}^{2} \equiv  m_{\nu_{\tau}}^{2} - m_{\nu_{\mu}}^{2} \simeq 
 \hat{m}_{\nu}^{2} \hat{\Delta}_{-}(2 + \hat{\Delta}_{-})$,  which leads to the 
approximate relation
\begin{eqnarray}
& &  \frac{\Delta m_{\mu e}^{2}}{\Delta m_{\tau\mu}^{2}}
\simeq  \left(\frac{\hat{\delta}_{-}}{\sqrt{2}\hat{\Delta}_{-}}\right)^{2} 
\simeq s^{2}_{\nu}= 2|U_{e3}|^{2}  << 1 \nonumber \\
& & \simeq  \begin{array}{ll}
0.2 \sim 0.09 &  $\quad$  MSW-LMA \\
0.02 \sim 0.002 & $\quad$  MSW-LOW \\
10^{-7} &  $\quad$ Vacuum\ Oscillation \\
\end{array} 
\end{eqnarray}
which implies that once the mixing element $|U_{e 3}|$ is determined, 
such a relation will tell us which solution should be favored.

  When going back to the weak gauge and charged-lepton mass basis, 
the neutrino mass matrix gets the following interesting form 
\begin{eqnarray}
& & M_{\nu}/\hat{m}_{\nu} \simeq \left( \begin{array}{ccc}
  0 & \frac{1}{\sqrt{2}}i & \frac{1}{\sqrt{2}}i  \\
   \frac{1}{\sqrt{2}}i & \frac{1}{2} &  - \frac{1}{2}  \\
  \frac{1}{\sqrt{2}}i &  - \frac{1}{2} &  \frac{1}{2}  \\ 
\end{array} \right)   \\
& & + \frac{\hat{\delta}_{-}}{2}\left( \begin{array}{ccc}
 0 & -\frac{1}{\sqrt{2}}i & \frac{1}{\sqrt{2}}i  \\
   -\frac{1}{\sqrt{2}}i & -1 &  0  \\
  \frac{1}{\sqrt{2}}i & 0 &  1 \\  
\end{array} \right) + \frac{\hat{\Delta}_{-}}{2}\left( \begin{array}{ccc}
 0 & 0 & 0  \\
   0 & 1 &  -1  \\
  0 & -1 &  1 \\  
\end{array} \right). \nonumber
\end{eqnarray}
As $(M_{\nu})_{ee} = 0$, the neutrinoless double beta decay
is forbiden in the model. Thus the neutrino masses can be approximately 
degenerate and large enough  ( $\hat{m}_{\nu}= O(1)$ eV) to play a 
significant cosmological role. Note that our scenario was shown to remain stable 
after considering renormalization group effects\cite{YLW}.

\section{Lepton Flavor Violations}

 The mass matrix of gauge fields $A_{\mu}^{i}$ is 
\begin{equation}
 M_{F}^{2} = \frac{m_{F}^{2}}{3}\left( \begin{array}{ccc}
  2(\xi_{+} + \xi')  & 0  & -\sqrt{2}\xi_{-} \\ 
    0 & 3\xi_{+} + \xi' & 0  \\ 
 -\sqrt{2}\xi_{-} & 0  &  3\xi_{+} + \xi'   \\ 
\end{array} \right)  
\end{equation} 
with $ m_{F}^{2}= 3g^{'2}_{3}\sigma^{2}/8$ and $\xi_{\pm}=(1 \pm \xi)/2$. 
This mass matrix is diagonalized 
by an orthogonal matrix $O_{F}$ via $O_{F}^{T}M_{F}^{2}O_{F}$ 
with $(O_{F})_{13}=\sin \theta_{F}\equiv  s_{F}$ and 
$\tan 2\theta_{F}  = 2\sqrt{2}\xi_{-}/(\xi_{+} - \xi')$. 
Denoting the physical gauge fields as $F_{\mu}^{i}$, 
we then have $A_{\mu}^{i} = O_{F}^{ij}F_{\mu}^{j}$. 
In the physical mass basis, we have for gauge interactions 
\begin{eqnarray}
{\cal L}_{F}  & = & \frac{g'_{3}}{2} F_{\mu}^{i}\bar{\nu}_{L}t^{j}
O_{F}^{ji}\gamma^{\mu}\nu_{L} \\
& + & \frac{g'_{3}}{2} F_{\mu}^{i}
\left( \bar{e}_{L} V_{e}^{i}\gamma^{\mu}e_{L} - \bar{e}_{R} V_{e}^{i \ast}
\gamma^{\mu}e_{R}\right) \nonumber 
\end{eqnarray}
with $V_{e}^{i} = U_{e}^{\dagger}t^{j}U_{e}O_{F}^{ji}$. Explicitly, we find 
\begin{eqnarray} 
V_{e}^{1} & = & \left( \begin{array}{ccc}
  c_{F} & i\frac{1}{2}s_{F} & -i\frac{1}{2}s_{F} \\
   -i\frac{1}{2}s_{F} & \frac{1}{2}c_{F} + \frac{1}{\sqrt{2}}s_{F} &  \frac{1}{2}c_{F}  \\
  i\frac{1}{2}s_{F} &  \frac{1}{2}c_{F} & \frac{1}{2}c_{F} -\frac{1}{\sqrt{2}}s_{F}  \\ 
\end{array} \right) \nonumber \\
 V_{e}^{2} & = & \left( \begin{array}{ccc}
  0 & \frac{1}{2} & -\frac{1}{2} \\
   \frac{1}{2} & 0 &  i\frac{1}{\sqrt{2}}  \\
 -\frac{1}{2} &  -i\frac{1}{\sqrt{2}} & 0 \\ 
\end{array} \right)  \\
V_{e}^{3} & = & \left( \begin{array}{ccc}
  -s_{F} & i\frac{1}{2}c_{F} & -i\frac{1}{2}c_{F} \\
   -i\frac{1}{2}c_{F} & -\frac{1}{2}s_{F} + \frac{1}{\sqrt{2}}c_{F} &  -\frac{1}{2}s_{F}  \\
  i\frac{1}{2}c_{F} &  -\frac{1}{2}s_{F} & -\frac{1}{2}s_{F} -\frac{1}{\sqrt{2}}c_{F}  \\ 
\end{array} \right).  \nonumber
\end{eqnarray} 
Thus the $SO(3)_{F}$ gauge interactions allow lepton flavor violating process 
$\mu \rightarrow 3e$, its branch ratio is 
\begin{equation}
Br(\mu \rightarrow 3e) = \left(\frac{v}{\sigma}\right)^{4}
 \frac{2\xi_{-}^{2}}{[(3\xi_{+} + \xi')(\xi_{+} + \xi') - \xi_{-}^{2} ]^{2} } 
\end{equation}
with $v=246$GeV. For $\sigma \sim 10^{3} v$, the branch ratio could be very close to 
the present experimental upper bound $Br(\mu \rightarrow 3e) < 1 \times 10^{-12}$. 
Thus when taking the mixing angle $\theta_{F}$ and the coupling constant $g'_{3}$ for the 
$SO(3)_{F}$ gauge bosons to be at the same order of magnitude as those for the 
electroweak gauge bosons, we find that masses of the SO(3)$_{F}$ gauge bosons 
are at the order of magnitudes $m_{F_{i}} \sim 10^{3} m_{W}\simeq 80 $ TeV. For smaller mising 
angle $\theta_F$, the SO(3)$_F$ gauge boson masses $m_{F_{i}}$ could be below 1 TeV.

\section*{Acknowledgments}
 This work was supported in part by Outstanding Young Scientist Research Fund 
under the grant No. 19625514.

\end{document}